\documentstyle[12pt,epsfig]{article}
\begin{document}
\thispagestyle{empty}
\hoffset=-1.2cm
\hsize=16cm
\vsize=24cm
\begin{flushright}
DTP-97/62\\
July 1997\\
\end{flushright}
\vskip 1cm
{\centerline{\bf {THE NON-PERTURBATIVE THREE-POINT VERTEX}}}
{\centerline{\bf {IN MASSLESS QUENCHED QED}}}
{\centerline{\bf {AND PERTURBATION THEORY CONSTRAINTS}}}
\vskip 1.5cm
\baselineskip=7mm
{\centerline{{\bf{{A. Bashir$^{\dagger}$, A. K{\i}z{\i}lers\"u$^{\dagger\dagger}$
 and M.R. Pennington$^{\dagger\dagger\dagger}$ }}}}
\vskip 5mm
{\centerline{$^{\dagger}$Department of Physics}}
{\centerline{Quaid-i-Azam University}}
{\centerline{Islamabad, Pakistan.}
\vskip 5mm
{\centerline{$^{\dagger\dagger}$Department of Physics}}
{\centerline{University of Istanbul}}
{\centerline{Istanbul, Turkey.}
\vskip 5mm
{\centerline{$^{\dagger\dagger\dagger}$Centre for Particle Theory}}
{\centerline{University of Durham}}
{\centerline{Durham DH1 3LE, U.K.}}
\vskip 1.2cm
{\centerline {ABSTRACT}}
\noindent
Dong, Munczek and Roberts~\cite{Dong} have shown how the full 3-point vertex
that appears in the Schwinger-Dyson equation for the
fermion propagator can be expressed in terms of a constrained function $W_1$ in
massless quenched QED. However, this analysis involved two key assumptions~:
that the fermion anomalous dimension vanishes in the Landau gauge and that the transverse
vertex has a simplified dependence on momenta. Here we remove these
assumptions and find the general form for a new constrained function $U_1$ that
ensures the multiplicative renormalizability of the fermion propagator non-perturbatively.
We then study the restriction imposed on $U_1$ by  recent perturbative
calculations of the vertex and compute its leading logarithmic expansion.
Since $U_1$ should
reduce to this expansion in the weak coupling regime, this
should serve as a guide to its non-perturbative construction. 
We comment on the perturbative
realization of the constraints on $U_1$.
\section{Introduction}
\baselineskip=8.mm
\parskip=2mm
\setcounter{page}{0}
The behaviour of the fermion propagator in any gauge theory is
determined by the fermion-gauge boson vertex. While in perturbation
theory, a bare vertex is sufficient, in the strong coupling regime it
is well-known that this simple {\em ansatz} can lead to unacceptable
results, such as  an unrenormalizable fermion propagator and
gauge-dependent chiral symmetry breaking.
The Ward-Green-Takahashi identity for the vertex determines what
is often called its longitudinal part~\cite{BC}. The remaining
transverse part has long been known to play a crucial role in ensuring
the multiplicative renormalizability of the fermion propagator
\cite{King,CP,Brown}.
However, it is only very recently that a general form for the
transverse vertex involving an odd number of gamma matrices has been
written down for quenched QED~\cite{Dong,BP1}. With simplifying assumptions,
this {\em ansatz} ensures that the
fermion propagator is multiplicatively renormalizable and that if a
dynamical mass is generated then this phase transition occurs at a
gauge-independent value of the critical coupling~\cite{BP1}.
This  {\em ansatz} involves two unknown
functions $W_1(x)$ and $W_2(x)$ of a dimensionless ratio $x$ of momenta
each satisfying an
integral and a derivative constraint. The integral condition on
$W_1$ guarantees that the fermion propagator is multiplicatively
renormalizable, whereas that on $W_2$ ensures that the critical coupling is a
gauge-independent quantity. The derivative
conditions are consequences of the transverse vertex being free of
kinematic singularities.
In the case of
massless fermions, $W_2$ drops out and only $W_1$ dictates what the
transverse vertex is. However this construction involves the assumption that
the transverse vertex vanishes in the Landau gauge and has no dependence on the
angle between the fermion momenta. Here we remove these assumptions
 and introduce
a new constrained function $U_1(x)$. In terms of this,
we present the most general non-perturbative construction of the transverse vertex
required by the multiplicative renormalizability of the fermion
propagator. 

\noindent In this paper,
we go on to discuss how perturbation theory
can provide additional constraints on $U_1$. Physically meaningful solutions
of the Schwinger-Dyson equations must agree with perturbation theory
in the weak coupling limit.
Its importance in dictating the non-perturbative structure of the
vertex has been appreciated in earlier work~\cite{CP}-\cite{Ayse}. 
We obtain the perturbative expansion of $U_1(x)$ to ${\cal O}(\alpha)$,
in the limit when $x \rightarrow 0$, to which every non-perturbative
construction
of $U_1$ must reduce.
This is made possible by the recent perturbative calculation of the
transverse vertex by K{\i}z{\i}lers\"u et al.~\cite{Ayse}
$U_1$ being related to a Green's function beyond lowest order,
is renormalization scheme dependent.
In this paper we have used the cut-off regularization scheme 
as is natural when discussing multiplicative renormalizability, whereas the
calculation of the transverse vertex by K{\i}z{\i}lers\"u  et al.~\cite {Ayse}
was performed in
the dimensional regularization scheme most useful in perturbation theory,
but which does not distinguish between ultraviolet and infrared behaviours. 
In order to retain consistency,
the perturbative evaluation of $U_1$ has been restricted to
leading logarithms alone, as these are  scheme-independent.
This fact, however, prevents us from checking explicitly that
the integral condition
on $U_1$ is preserved in perturbation theory, though  consistency requires it is.

\noindent We check the validity of the derivative constraint on $U_1$ in perturbation theory.
This condition holds in the limit when $x \rightarrow 1$. Analytical
calculation of $U_1$ and its  derivative in this region is a prohibitively
difficult task. However, numerical evaluation is possible. We find that
the  numerical results  are  in excellent  agreement  with the proposed
condition.


\section{Wavefunction Renormalization F($p^2,\Lambda^2$)}
\baselineskip=8.5mm

The Schwinger-Dyson equation for the fermion propagator, $S_{F}(p)$,
in QED with a bare
coupling, $e$, is displayed in Fig 1, and is given by~:
\begin{eqnarray}
     iS_{F}^{-1}(p)\,=\,iS_{F}^{0^{-1}}(p)\,-\,e^2\int \frac{d^4k}{(2\pi)^4}\,
    \gamma^{\mu}\,S_{F}(k)\, \Gamma^{\nu}(k,p)\,\Delta_{\mu\nu}
    (q) \quad , \label{eq:SDFP}
\end{eqnarray}
where  $q=k-p$. For massless fermions, $S_{F}(p)$ can be
expressed in terms of a single Lorentz scalar
function, $F(p^2,\Lambda^2)$, called the wavefunction renormalization,
so that
\begin{eqnarray*}
    S_{F}(p)&=& \frac{F(p^2,\Lambda^2)}{\not \! p}\qquad ,
\end{eqnarray*}
where $\Lambda$ is the ultraviolet cut-off used to indicate that
the integrals involved are divergent and need to be regularized.
The bare propagator $ S_{F}^{0}(p)= 1/ \rlap / p $. The photon
propagator remains unrenormalized in quenched QED~:
\begin{eqnarray*}
 \Delta_{\mu\nu}(q)\, =\,\frac{1}{q^2}\ \left(
    g_{\mu\nu}+(\xi-1)\frac{q_{\mu}q_{\nu}}{q^2}\ \ \right) \equiv
     \Delta_{\mu\nu}^{T}(q)+\xi\frac{q_{\mu}q_{\nu}}{q^4} \qquad ,
\end{eqnarray*}
where $\Delta_{\mu\nu}^{T}(q)$, called the transverse part of the
propagator, is defined by the above equation
and $\xi$ is the standard covariant gauge parameter.
$\Gamma^{\mu}(k,p)$ is the full fermion-boson vertex, for which we must
make an {\em ansatz} in order to solve Eq.~(\ref{eq:SDFP}).
Keeping in mind that the vertex satisfies the Ward-Green-Takahashi
identity
\begin{eqnarray}
    q^{\mu}\Gamma_{\mu}(k,p)= S_{F}^{-1}(k)-S_{F}^{-1}(p) \quad,
\label{eq:WTI}
\end{eqnarray}
Ball and Chiu~\cite{BC} considered the vertex as a sum of longitudinal
and transverse components~:
 \begin{eqnarray}
      \Gamma^{\mu}(k,p)=\Gamma^{\mu}_{L}(k,p)+\Gamma^{\mu}_{T}(k,p)\quad ,
\label{eq:decomp}
\end{eqnarray}
where $\Gamma^{\mu}_{T}(k,p)$ is defined by~:
\begin{eqnarray}
    q_{\mu} \Gamma^{\mu}_{T}(k,p)=0 \quad .  \label{eq:DTV}
\end{eqnarray}
To satisfy Eq.~(\ref{eq:SDFP}) in a manner free of kinematic singularities,
which in turn ensures the Ward identity is fulfilled, we have
(following Ball and Chiu)~:
\begin{eqnarray}
\Gamma^{\mu}_{L}(k,p)= a(k^2,p^2) \gamma^{\mu}
              + \mbox{} b(k^2,p^2) (\not \! k + \not \! p)
               (k+p)^{\mu} \quad,  \label{eq:BCV}
\end{eqnarray}
 where
\begin{eqnarray}
\nonumber
  a(k^2,p^2)&=&\frac{1}{2}\ \left( \frac{1}{F(k^2,\Lambda^2)}\ +
  \frac{1}{F(p^2,\Lambda^2)}\  \right) \qquad \qquad , \\
  b(k^2,p^2)&=&\frac{1}{2}\ \left(
  \frac{1}{F(k^2,\Lambda^2)} - \frac{1}{F(p^2,\Lambda^2)}\ \right)
\frac{1}{k^2 - p^2}\quad    \label{eq:BCAB}
\end{eqnarray}
and
\begin{eqnarray}
       \Gamma^{\mu}_{T}(p,p)=0 \quad . \label{eq:WI}
\end{eqnarray}
Ball and Chiu~\cite{BC} demonstrated that a set of 8 vectors $T_{i}^{\mu}(k,p)$
formed a general basis
for the transverse part, so that~:
\vspace {2mm}
\begin{eqnarray}
   \Gamma^{\mu}_{T}(k,p)= \sum_{i=1}^8 \tau_{i}(k^2,p^2,q^2)\ T_{i}^
     {\mu}(k,p)    \label{eq:basis}
\end{eqnarray}
Eqs.~(\ref{eq:DTV},\ref{eq:WI}) are then satisfied
provided that in the limit $k\rightarrow p $, the $\tau_{i}(p^2,p^2,0)$
are finite. As shown by K{\i}z{\i}lers\"u et al.~\cite{Ayse}
a modification of the original Ball-Chiu basis is required to achieve this
in an arbitrary covariant gauge in perturbation theory. One can then define
the Minkowski space basis to be~\cite{Ayse}~:
\vspace{2 mm}
\begin{eqnarray}
 \nonumber  T_{1}^{\mu}(k,p)&=& p^{\mu}(k.q)-k^{\mu}(p.q)                        \\
 \nonumber   T_{2}^{\mu}(k,p)&=& T_{1}^{\mu} ( \not \! k +\not \! p)              \\
 \nonumber   T_{3}^{\mu}(k,p)&=& q^{2} \gamma^{\mu}-q^{\mu} \not \! q             \\
 \nonumber   T_{4}^{\mu}(k,p)&=& q^2 [ \gamma^{\mu}
            ( \not \! k +\not \! p ) - k^{\mu} - p^{\mu}]
               - 2(k-p)^{\mu} \sigma_{\lambda\nu}  k^{\lambda} p^{\nu}   \\
 \nonumber   T_{5}^{\mu}(k,p)&=& -\sigma^{\mu\nu} q_{\nu}                          \\
 \nonumber   T_{6}^{\mu}(k,p)&=& \gamma^{\mu} (k^{2} - p^{2})
             -( k+p)^{\mu} ( \not \! k -\not \! p )         \\
 \nonumber   T_{7}^{\mu}(k,p)&=& -\frac{1}{2} (k^{2} - p^{2}) [ \gamma^{\mu}
            ( \not \! k +\not \! p ) - k^{\mu} - p^{\mu}]
               +(k+p)^{\mu} \sigma_{\lambda\nu}  k^{\lambda} p^{\nu}  \\
  T_{8}^{\mu}(k,p)&=&  \gamma^{\mu} \sigma_{\lambda\nu} k^{\lambda} p^{\nu}  
               - k^{\mu} \not \! p + p^{\mu} \not \! k   \
\qquad.   \label{eq:basis1}
\end{eqnarray}
On multiplying Eq.~(\ref{eq:SDFP}) by  $\not \! p$, taking
the trace and making use of
Eqs.~(\ref{eq:WTI},\ref{eq:decomp},\ref{eq:BCV},\ref{eq:BCAB},\ref{eq:basis},\ref{eq:basis1}) we have on Wick rotating to Euclidean space~:
\begin{eqnarray}
\nonumber
 \frac{1}{F(p^2,\Lambda^2)}\,=\; 1\;-
             &&\frac{\alpha}{4\pi^3} \frac{1}{p^2} \int
              d^4k \,
              \frac{F(k^2,\Lambda^2)}{k^2q^2}\
                \\   \nonumber   \
           & &  \Bigg\{ \hspace{16 mm}\vspace{5 mm} a(k^2,p^2)
               \frac{1}{q^2}
             \left[ -2 \Delta^2 -3q^2 k\cdot  p \right]
              \\  \nonumber
           & &+ \hspace{16 mm}b(k^2,p^2) \frac{1}{q^2}
            \left[ -2 \Delta^2 (k^2+p^2)
            \right]  \\  \nonumber
           & &-\hspace{16mm} \frac{\xi} {F(p^2,\Lambda^2)}\ \frac{p^2}{q^2}
               \ (k^2-k\cdot p)  \\ \nonumber
           & &+\hspace{16mm}\tau_{2}(k^2,p^2,q^2)
               \left[- \Delta^2  (k^2+p^2) \right]   \\ \nonumber
            \nonumber
           & &+\hspace{16mm}\tau_{3}(k^2,p^2,q^2)
               \left[ 2 \Delta^2 + 3 q^2 k\cdot p  \right]
             \\  \nonumber
           & &+\hspace{16 mm}\tau_{6}(k^2,p^2,q^2)
               \left[3(k^2-p^2)k\cdot p  \right] \\
           & &+\hspace{16 mm}\tau_{8}(k^2,p^2,q^2)
               \left[ 2 \Delta^2 \right] \Bigg\} \quad. \label{eq:FEX}
\end{eqnarray}
where $\Delta^2 = (k \cdot p)^2 - k^2 p^2 $. Note that only those
$T_i^\mu$ with odd numbers of gamma matrices contribute in the case of
massless fermions --- incidentally, these are then the  same as in the
basis proposed in~\cite{BC}. At this stage,
it appears impossible to proceed any further without demanding
that the $\tau_i$ are independent of the angle between the 
fermion momentum vectors $k$ and $p$, {\it i.e.} independent of
$q^2$. This assumption allows us to carry out the angular integration
in Eq.~(\ref{eq:FEX}).
We shall show later in this paper that
this assumption is not a necessary requirement for solving the above
Schwinger-Dyson equation and this can readily be
undone. In order to distinguish the transverse components which are
assumed to be independent of $q^2$ from the real ones which
explicitly depend on $q^2$~\cite{Ayse}, we denote the
former by $\tau_i^{\rm{eff}}$, suggesting that these are only effective
$\tau_i$. Now carrying out the angular integration~:
\vspace{2mm}
\begin{eqnarray}
 \nonumber
 \frac{1}{F(p^2,\Lambda^2)}\,&=&\; 1\; - \frac{\alpha}{4\pi}
           \int_{0}^{\Lambda^2}  \frac{d k^2}{k^2} \;
              \, F(k^2,\Lambda^2)\
                \\   \nonumber
           & &  \Bigg[ \hspace{2 mm} \frac{k^{4}}{p^{4}}
             \hspace{2 mm}  \Bigg\{ \hspace{2 mm} b(k^2,p^2)
             \left[ \frac{3}{2} (k^2 + p^2) \right]
              \\  \nonumber
           & & \hspace{10 mm} + \hspace{14 mm}\tau_{2}^{{\rm
               eff}}(k^2,p^2)
               \left[- \frac{1}{4} (k^2+p^2) (k^2-3p^2) \right] \\
            \nonumber
           & & \hspace{10 mm} + \hspace{14 mm}\tau_{3}^{{\rm
            eff}}(k^2,p^2)
               \left[ \frac{1}{2} (k^2-3p^2)  \right]
             \\
           \nonumber
           & & \hspace{10 mm} + \hspace{14 mm}\tau_{6}^{{\rm
            eff}}(k^2,p^2)
               \left[ \frac{3}{2} (k^2-p^2) \right] \\
  \nonumber        & & \hspace{10 mm} + \hspace{14 mm}\tau_{8}^{{\rm
            eff}}(k^2,p^2)
               \left[ \frac{1}{2} (k^2-3p^2) \right] \Bigg\}
            \theta(p^2 - k^2)  \\
  \nonumber        & &   \hspace{6 mm} + \hspace{2 mm} \Bigg\{
              \hspace{2 mm} b(k^2,p^2)
             \left[ \frac{3}{2} (k^2 - p^2) \right]
             - \frac{\xi}{F(p^2)}\ \\
            \nonumber
           & & \hspace{10 mm} + \hspace{14 mm}\tau_{2}^{{\rm
              eff}}(k^2,p^2)
               \left[- \frac{1}{4} (k^2+p^2) (p^2-3k^2) \right] \\
            \nonumber
           & & \hspace{10 mm} + \hspace{14 mm}\tau_{3}^{{\rm
              eff}}(k^2,p^2)
               \left[ \frac{1}{2} (p^2-3k^2)  \right]
             \\
           \nonumber
           & & \hspace{10 mm} + \hspace{14 mm}\tau_{6}^{{\rm
              eff}}(k^2,p^2)
               \left[ \frac{3}{2} (k^2-p^2) \right] \\
           & & \hspace{10 mm} + \hspace{14 mm}\tau_{8}^{{\rm
              eff}}(k^2,p^2)
               \left[ \frac{1}{2} (p^2-3k^2) \right] \Bigg\}
            \theta(k^2 - p^2)             \Bigg] \,. \label{eq:FEF}
\end{eqnarray}
Following Dong, Munczek and Roberts~\cite{Dong},
 Bashir and Pennington~\cite{BP1}  have proposed an {\em ansatz}
for the transverse vertex. They require it to
be chosen such that the fermion propagator is multiplicatively
renormalizable and in the case of massive fermions, the chiral symmetry
breaking phase transition takes place at a gauge-invariant value of
the coupling. They show that the transverse vertex can be written in
terms of two unknown functions $W_1$ and $W_2$, each obeying an
integral and a derivative condition. In the case of a chirally symmetric
solution, the transverse vertex reduces to being a function of $W_1$
alone. However, this construction involves the additional assumption that the transverse vertex is zero in the Landau gauge.
In general, the solution of Eq.~(11) imposed by
multiplicative renormalizability in quenched QED is 
\begin{eqnarray}
F(p^2, \Lambda^2)\,=\, A\ \left(p^2/\Lambda^2\right)^{\gamma}
\end{eqnarray}
in any covariant gauge with $A$ a constant. The assumption of a vanishing transverse vertex in the
Landau gauge means that the anomalous dimension, $\gamma$ is equal to
$\nu \equiv \alpha \xi/4\pi$. Crucially this is not the general solution,
nor is it even in agreement with perturbation theory~\cite{Ross}. The anomalous dimension,
$\gamma$, is not  zero in the Landau gauge.  Consequently,
we fix the effective transverse vertex quite generally
in terms of a function $U_1(x)$ through a series of steps analogous to those followed
in Refs.~\cite{Dong,BP1}.

\noindent The result is~:
\begin{eqnarray}
\nonumber
   \overline{\tau}_{\rm{eff}}(k^2,p^2)\;=
     && \;\ \frac{1}{4}\ \frac{1}{k^2-p^2}\
    \frac{1}{s_{1}(k^2,p^2)}\  \Bigg[
     U_{1}\left( \frac{k^2}{p^2}\ \right) - U_{1}\left(\frac{p^2}{k^2}\
     \right) \Bigg]\\ 
     && - \frac{2\pi}{\alpha}\ \frac{\gamma-\nu}{k^2-p^2}
    \left( \frac{1}{F(k^2, \Lambda^2)} - \frac{1}{F(p^2,\Lambda^2)} \right)
       \label{eq:tbar}
    \\ \nonumber
   \\ \nonumber
\tau_{6}^{\rm{eff}}(k^2,p^2)\;=&&-\,\frac{1}{2}\ \frac{k^2+p^2}{(k^2-p^2)^2}
    \left( \frac{1}{F(k^2, \Lambda^2)} - \frac{1}{F(p^2,\Lambda^2)} \right)      +
      \frac{1}{3}\, \frac{k^2+p^2}{k^2-p^2}\
    \overline{\tau}_{\rm{eff}}(k^2,p^2) \\ \nonumber
  &&+ \,\frac{1}{6}\ \frac{1}{k^2-p^2}\  \frac{1}{s_{1}(k^2,p^2)}\
     \Bigg[ U_{1}\left(\frac{k^2}{p^2}\ \right)  +
    U_{1}\left(\frac{p^2}{k^2}\ \right)  \Bigg]\\ 
    &&  - \frac{4\pi}{3\alpha}\ \frac{\gamma-\nu}{k^2-p^2}
    \left( \frac{1}{F(k^2, \Lambda^2)} + \frac{1}{F(p^2,\Lambda^2)} \right)\label{eq:t6}
\end{eqnarray}
where
\begin{eqnarray*}
 s_{1}(k^2,p^2)= \frac{k^2}{p^2}\ F(k^2,\Lambda^2) +
\frac{p^2}{k^2}\ F(p^2,\Lambda^2)
\end{eqnarray*}
and
\begin{eqnarray}
    \overline{\tau}_{\rm{eff}}(k^2,p^2)
    &=&\tau_{3}^{\rm{eff}}(k^2,p^2) + \tau_{8}^{\rm{eff}}(k^2,p^2) -
          \frac{1}{2}\ (k^2+p^2)\, \tau_{2}^{\rm{eff}}(k^2,p^2)\quad .
\end{eqnarray}

\noindent To see how these forms arise uniquely, let us substitute Eqs.~(13,14)
into Eq.~(11) to obtain
\begin{eqnarray}
\nonumber
 \frac{1}{F(p^2,\Lambda^2)} &=& 1+ \nu
\int_{p^2}^{\Lambda^2} \frac{dk^2}{k^2}
\frac{F(k^2,\Lambda^2)}{F(p^2,\Lambda^2)}  \\ \nonumber
&& \hspace{3mm} - \frac{\alpha}{8 \pi} \int_{0}^{p^2} \frac{dk^2\ k^2}{p^4}
\frac{ U_1\left(k^2/p^2\right) F(k^2,\Lambda^2)}{\left(k^2/p^2\right)
F(k^2,\Lambda^2)
+  \left(p^2/k^2\right)  F(p^2,\Lambda^2)}  \\ \nonumber
&& \hspace{3mm} - \frac{\alpha}{8 \pi} \int_{p^2}^{\Lambda^2}
\frac{dk^2}{k^2} \hspace{4mm}  \frac{ U_1\left(p^2/k^2\right)
F(k^2,\Lambda^2)}{  \left(k^2/p^2\right) F(k^2,\Lambda^2)
+  \left(p^2/k^2\right) F(p^2,\Lambda^2)}\\ 
&& \hspace{3mm} + (\gamma - \nu) \int_0^{p^2} \frac{dk^2\ k^2}{p^4}\hspace{4mm}
+ \hspace{4mm} (\gamma - \nu) \int_{p^2}^{\Lambda^2} 
\frac{dk^2}{k^2}
\frac{F(k^2,\Lambda^2)}{F(p^2,\Lambda^2)} \;. \label{eq:GFP}
\end{eqnarray}
where recall $\nu \equiv \alpha\xi/(4\pi)$. Multiplicative
renormalizability requires that
the renormalized  $F_{R}$  is related
to the unrenormalized $F$ through a multiplicative factor $Z$ by~:
\begin{eqnarray}
 F(p^2,\Lambda^2) = Z\left(\mu^2/\Lambda^2\right) F_{R}(p^2,\mu^2) \quad.
\end{eqnarray}
so that the solution of this equation is
\begin{eqnarray}
\frac {F_R(k^2,\mu^2)}{F_R(p^2,\mu^2)}\,=\,\frac{F(k^2,\Lambda^2)}{F(p^2,\Lambda^2)}
\,=\,\left(\frac{k^2}{p^2}\right)^{\gamma}\; .
\end{eqnarray}
Now this power behaviour is the solution of 
\begin{eqnarray}
 \frac{1}{F(p^2,\Lambda^2)} &=& 1+ \gamma
\int_{p^2}^{\Lambda^2} \frac{dk^2}{k^2}
\frac{F(k^2,\Lambda^2)}{F(p^2,\Lambda^2)} \; .
\end{eqnarray}
Consequently, from Eq.~(16), this imposes the following restriction on the transverse vertex 
and hence the function $U_1(k^2/p^2)$~:
\begin{eqnarray}
\nonumber &&
\frac{\alpha}{8 \pi} \int_{0}^{p^2} \frac{dk^2 k^2}{p^4}
\frac{ U_1\left(k^2/p^2\right) \left(F(k^2,\Lambda^2)/F(p^2,\Lambda^2)\right)}{ \left(k^2/p^2\right)
\left(F(k^2,\Lambda^2)/F(p^2,\Lambda^2)\right)
+  \left(p^2/k^2\right) }  \\
&&  + \frac{\alpha}{8 \pi} \int_{p^2}^{\Lambda^2}
\frac{dk^2}{k^2} \hspace{5mm}  \frac{ U_1\left(p^2/k^2\right)}{ { \left(k^2/p^2\right)}
+\left(p^2/k^2\right) \left(F(p^2,\Lambda^2)/F(k^2,\Lambda^2)\right) }
- \frac {1}{2} (\gamma - \nu) \,=\, 0
\label{eq:inter1} \; .
\end{eqnarray}
Introducing the variable $x$, where
\begin{eqnarray}
\nonumber
 && x=k^2/p^2   \hspace{10mm}  \forall \hspace{5mm}   0 \le k^2 < p^2 \\
 && x=p^2/k^2   \hspace{10mm}  \forall \hspace{5mm}   p^2 \le k^2 < \Lambda^2
\label{eq:defx}
\end{eqnarray}
in the first two terms of the above equation, Eq.~(20) becomes simply
\begin{eqnarray}
\int_0^1\, dx\ \frac{U_1(x) \ x^{1+\gamma}}{x^{-1}\,+\,x^{1+\gamma}}\,+\,
\int^1_{p^2/\Lambda^2}\, dx\ \frac{U_1(x) \ x^{-1}}{x^{-1}\,+\,x^{1+\gamma}}\,
=\, \frac{4\pi}{\alpha}\ (\gamma - \nu)\, .
\end{eqnarray}
We can now let $\Lambda^2 \to  \infty$ and so we simply have
\begin{eqnarray}
\int_0^1\, dx\ U_1(x)\,=\, \frac{4\pi}{\alpha} (\gamma - \nu) \; .
\end{eqnarray}
Note that the previous construction~\cite{Dong,BP1} explicitly assumed that $\gamma =
\nu =\alpha \xi/(4\pi)$ and then $U_1(x) \to W_1(x)$ and $\int_0^1 dx W_1(x) = 0$.
Moreover, the simplified vertex of \cite{CP} corresponds to setting $W_1(x)=0$.

\noindent The transverse vertex has no kinematic singularities. Motivated
by the perturbative calculation of Ball and Chiu~\cite{BC} in the Feynman gauge
and later by K{\i}z{\i}lers\"u  et al.~\cite{Ayse} 
in arbitrary covariant gauges, it is a plausible assumption
that even non-perturbatively this is achieved by 
the individual
$\tau_i$'s being free of kinematic singularities. The antisymmetry
of $\tau^{\rm{eff}}_6(k^2,p^2)$ under $k^2 \leftrightarrow p^2$ interchange 
then requires that
\begin{eqnarray}
\lim_{k^2\to p^2}\ (k^2 - p^2)\ \tau^{\rm{eff}}_6(k^2,p^2)\,=\,0\, .
\end{eqnarray}
This imposes another constraint on $U_1(x)$~:
\begin{eqnarray}
U_1(1)\,+\,U_1^{\prime}(1)\,=\, -6 \gamma\,+\,
\frac{8\pi}{\alpha} (\gamma - \nu) (2- \gamma)\, .
\end{eqnarray}

\noindent For later, let us note that Eqs.~(\ref{eq:tbar}) and (\ref{eq:t6}) can be inverted to write $U_1$
in terms of the $\tau_i^{\rm{eff}}$~:
\begin{eqnarray}
\nonumber
U_1 \left( \frac{k^2}{p^2} \right) &=& s_1(k^2,p^2) \Bigg[ (k^2-3p^2)
\overline{\tau}_{\rm{eff}}(k^2,p^2) + \frac{3}{2} \frac{k^2+p^2}{k^2-p^2} \left(
\frac{1}{F(k^2,\Lambda^2)} - \frac{1}{F(p^2,\Lambda^2)} \right) \\
&& \hspace{2.3cm} + 3 (k^2 - p^2) \tau_6^{\rm{eff}}
(k^2,p^2) + \frac{8\pi}{\alpha} (\gamma - \nu) \frac{1}{F(k^2,\Lambda^2)}  \Bigg] \, . \label{eq:DW1}
\end{eqnarray}
We now set about  using recent perturbative calculations
of the structure of the vertex to determine  the weak coupling limit of this $U_1$.
\newpage
\section{Real Vertex and Effective Vertex}
\baselineskip=9mm

Calculation of $U_1(x)$ is
non-trivial. This is because
to be able to
solve the Schwinger-Dyson equation for the fermion propagator requires
assumptions to be made about the way the fermion-boson vertex
$\Gamma^{\mu}(k,p,q)$ depends upon~$q^2$. Indeed, it seems impossible to 
proceed analytically without assuming the vertex is independent of the photon
momentum $q$, otherwise
we cannot carry out the integration over the angular
variable. A motivation for this simplifying
assumption comes from the large momentum
behaviour of the vertex in perturbation theory, where it does, indeed, only depend on the
variables $k^2$ and $p^2$, and {\bf{ not on $q^2$}}~\cite{CP}~:
\vspace {2mm}
\begin{eqnarray}
\Gamma_T^{\mu}(k,p)\,\simeq\,-\, \frac {\alpha\xi}{8\pi} \
                             \,\ln{k^2\over{p^2}}\,
\left[ \gamma^{\mu}\,-\,{k^{\mu} \rlap / k \over{k^2}} \right]\quad .
 \label{eq:largek}
\end{eqnarray}
However, it is
clear from the perturbative calculation of K{\i}z{\i}lers\"u, Reenders and
Pennington~\cite{Ayse} that the same does not hold true for all the ranges of
$k^2$ and $p^2$. Instead, the $q^2$-dependence occurs in almost every
term of each of the $\tau_{i}$. We should, therefore,  keep in mind that
whenever we are neglecting the $q^2$-dependence, we are not talking about
the exact but only the {\it effective} vertex. In order to
find a connection between the two, we compare Eqs.~(\ref{eq:FEX}) and
(\ref{eq:FEF}), which yields the following exact
relation between the real, and the effective, $\tau_{i}$~:
\vspace{2mm}
\begin{eqnarray}
\nonumber
 \tau_2^{\rm eff}(k^2,p^2)&=& \frac{1}{f(k^2,p^2)}\int_{0}^{\pi} d \theta \;
\frac{\sin^2 \theta}{q^2}\;
\tau_2(k^2,p^2,q^2) \;  \Delta^2   \\
\nonumber
 \tau_3^{\rm eff}(k^2,p^2)&=&  \frac{1}{f(k^2,p^2)}\int_{0}^{\pi} d \theta\;
\frac{\sin^2 \theta}{q^2}\;
\tau_3(k^2,p^2,q^2) \left( \Delta^2 + \frac{3}{2} q^2 k\cdot p \right) \\
\nonumber
 \tau_6^{\rm eff}(k^2,p^2)&=& \frac{1}{f_{6}(k^2,p^2)}\int_{0}^{\pi} d \theta\;
  \frac{\sin^2 \theta}{q^2}
\;\tau_6(k^2,p^2,q^2)\;  k\cdot p  \\
\nonumber
 \tau_8^{\rm eff}(k^2,p^2)&=&  \frac{1}{f(k^2,p^2)}\int_{0}^{\pi} d \theta\;
\frac{\sin^2 \theta}{q^2}\;
\tau_8(k^2,p^2,q^2) \; \Delta^2 \quad, \\ \label{eq:REAE}
\end{eqnarray}
where
\begin{eqnarray}
\nonumber
f(k^2,p^2)&=& \frac{\pi}{8} \; \Bigg[ \; \frac{k^2}{p^2} (k^2-3p^2) \; \theta(p^2-k^2)+
\frac{p^2}{k^2} (p^2-3k^2) \; \theta(k^2-p^2) \; \Bigg]  \\
\nonumber
f_6(k^2,p^2)&=& \frac{\pi}{4} \; \Bigg[  \;\frac{k^2}{p^2}  \;
\theta(p^2-k^2)+
\frac{p^2}{k^2}  \; \theta(k^2-p^2)  \; \Bigg] \quad.
\end{eqnarray}
The perturbative evaluation of $\tau_i^{\rm{eff}}$ using Eq.~(\ref{eq:REAE})
is made possible by the calculation of K{\i}z{\i}lers\"u  et al.~\cite{Ayse} for the real
$\tau_i$~:
\vspace{2mm}
\begin{eqnarray}
\nonumber
 \tau_2(k^2,p^2,q^2)&=&\frac{\alpha}{8\pi \Delta^2} \hspace{2 mm}
           \Bigg\{ \hspace{2 mm} J_0 \hspace{5.6 mm}
          \Bigg[\, {1\over 2}(\xi-2) \left(
\frac{3}{2 \Delta^2} q^2 k^2 p^2 + (k^2 + p^2) \right) +  k\cdot  p\,
\Bigg]  \\
\nonumber
&& \hspace{12 mm} - {\rm ln} \frac{k^2}{p^2} \hspace{3.6 mm} \Bigg[\,
(\xi-2) \, \frac{3}{4 \Delta^2} (k^2-p^2) k\cdot  p +
\frac{\xi}{2} \frac{ (k+p)^2}{k^2 -p^2}
\Bigg]  \\
\nonumber
&& \hspace{12 mm} + {\rm ln} \frac{q^4}{k^2 p^2}  \Bigg[
(\xi-2) \,
\frac{3}{4 \Delta^2} \, q^2 \, k\cdot  p + \xi -1  \Bigg]  \\
&& \hspace{12 mm} + \hspace{14.5 mm}   (\xi-2) \;
\Bigg\} \label{eq:Ayse2}  \\  \nonumber \\
\nonumber
 \tau_3(k^2,p^2,q^2)&=&\frac{\alpha}{8\pi \Delta^2} \hspace{2 mm}
          \Bigg\{ \hspace{2 mm}  J_0 \hspace{5.6 mm}
          \Bigg[ \frac{(\xi-2)}{8} \left( 
\frac{3}{ \Delta^2} (k^2 - p^2)^2  ( k\cdot  p )^2 + (k^2 + p^2)^2
\right) -   \Delta^2 \Bigg]  \\
\nonumber
&& \hspace{12 mm} + {\rm ln} \frac{k^2}{p^2} \hspace{3.6 mm} \Bigg[
(\xi-2) \,  \frac{k^2 -p^2}{4} \left( 1 - \frac{3}{2 \Delta^2} (k+p)^2
k\cdot  p \right) \Bigg]  \\
\nonumber
&& \hspace{12 mm} + {\rm ln} \frac{q^4}{k^2 p^2}  \Bigg[ (\xi-2) \,
\frac{k\cdot  p}{2} \left(
\frac{3}{4 \Delta^2}  (k^2-p^2)^2 -1 \right)  \Bigg]  \\
&& \hspace{12 mm} + \hspace{11 mm} {1\over 2}\ (\xi-2) \,
(k+p)^2
\Bigg\} \label{eq:Ayse3} \\ \nonumber \\
\nonumber
 \frac{\tau_6(k^2,p^2,q^2)}{k^2-p^2}&=&
 \frac{\alpha (\xi-2)}{32\pi \Delta^2}
        \Bigg\{ \hspace{2 mm} J_0 \hspace{5.6 mm}
          \Bigg[  \frac{q^2}{2} \left( 1 -
\frac{3}{ \Delta^2} ( k\cdot  p )^2 \right) +   \Delta^2  \Bigg]
  \\
\nonumber
&& \hspace{12 mm} + {\rm ln} \frac{k^2}{p^2} \hspace{3.6 mm} \Bigg[
 \frac{3}{2 \Delta^2} k\cdot p (k^2-p^2) - \frac{(k+p)^2}{k^2-p^2}
 \Bigg]  \\
&& \hspace{12 mm} + {\rm ln} \frac{q^4}{k^2 p^2}  \Bigg[
\frac{-3}{2 \Delta^2} \, q^2 \, k\cdot  p  \Bigg]\; - \; 2 \;
\Bigg\}  \label{eq:Ayse6}  \\  \nonumber \\
\tau_8(k^2,p^2,q^2)&=&\frac{\alpha}{8\pi \Delta^2} \hspace{2 mm} \Bigg\{ q^2 \, \Bigg[
k\cdot p J_0 +  \, {\rm ln} \frac{q^4}{k^2 p^2} \Bigg] - (k^2-p^2)  \, {\rm ln}
\frac{k^2}{p^2} \Bigg\} \, , \label{eq:Ayse8} \\ \nonumber
\end{eqnarray}
where
\begin{eqnarray}
\nonumber
  J_0 &=& \frac {2}{ i \pi^2}  \int d^4 \omega \; \frac{1}{ \omega^2
( \omega - p)^2 ( \omega - k)^2} \quad , \\  \nonumber \\
 &=& \frac{2}{\Delta} \Bigg[ f \left( \frac{ k\cdot p -
\Delta}{p^2} \right) - f \left( \frac{ k\cdot p + \Delta}{p^2} \right)
+ \frac{1}{2} \; {\rm ln} \frac{q^2}{p^2} \; {\rm ln} \left( \frac{ k\cdot p -
\Delta}{k\cdot p + \Delta} \right) \Bigg] \quad, 
\end{eqnarray}
and
\begin{eqnarray}
 f(x) = {\rm Sp}(1-x) \;=\; - \int_{x}^{1} dy \, \frac{ {\rm ln}y}{1-y} \qquad.
\end{eqnarray}
Although the
Eqs. (\ref{eq:Ayse2}-\ref{eq:Ayse8})
appear a little complicated, the nice thing is that
all the $\tau_{i}$ are expressed in terms of elementary functions and
a single scalar integral $J_{0}$.
K{\i}z{\i}lers\"u  et al. have carried out the calculation in the
dimensional regularization scheme, whereas here, so as to be able to identify
the ultraviolet
behaviour readily, we use the cut-off method.
Consequently, we restrict our
discussion to  leading logarithms, which are independent of the choice of the
regularization scheme.
In the asymptotic limit $k^2 \gg p^2$, the integrals can be evaluated
analytically and the separation between the leading and the next to
leading terms becomes apparent.

In order to have a perturbative expansion for $U_1$, we have to go
up to ${\cal O}(1/k^{4})$ in $\tau^{\rm eff}_3$, $\tau^{\rm eff}_6$ and
$\tau^{\rm eff}_8$, and ${\cal O}(1/k^{6})$
in $\tau^{\rm eff}_2$,  instead of just keeping the terms of order ${\cal
O}(1/k^{2})$ and ${\cal
O}(1/k^{4})$ respectively.
Consequently, in an
arbitrary gauge, we have to go up to ${\cal O}(1/k^{7})$ in evaluating
$J_0$ for $k^2$ large. The expansion of $J_0$, keeping only the
logarithms,
to the required order in the limit when $k^2 \gg
p^2$ is~:
\vspace{2 mm}
\begin{eqnarray}
\nonumber
J_0&=& \frac{2}{k^2} \Bigg[  1+ \frac{ k\cdot p}{k^2} -
\frac{1}{3} \frac{p^2}{k^2} + \frac{4}{3} \frac{ ( k\cdot p )^2}{k^4}
- \frac{p^2 k \cdot p}{k^4} +2 \frac{(k \cdot p)^3}{k^6} + \frac{1}{5}
\frac{p^4}{k^4} -\frac{12}{5} \frac{p^2 (k \cdot p)^2}{k^6}\\
\nonumber
&&\hspace{8 Mm}
+ \frac{16}{5} \frac{(k \cdot p)^4}{k^8} + \frac{p^4 k \cdot p}{k^6} -
\frac{16}{3} \frac{p^2 (k \cdot p)^3}{k^8} + \frac{16}{3} \frac{(k \cdot
p)^5}{k^{10}}  \Bigg] \quad {\rm ln} \frac{k^2}{p^2} \; .\\
\end{eqnarray}
Now the perturbative expansion of the real $\tau_i$ can
be written as~:
\begin{eqnarray}
\nonumber
\tau_2(k^2,p^2,q^2)&=&-\frac{\alpha}{12 \pi k^4} \Bigg\{ \hspace{5mm}
1 + \, \, 2 \, \, \frac{k
\cdot p}{k^2} + \, \frac{1}{5k^4} \hspace{5mm}\left( 18 (k \cdot p)^2 -k^2 p^2
\right) \\
\nonumber
&& \hspace{15mm} - \xi \Bigg[ 2 + \, \, 3 \, \, \frac{k \cdot p}{k^2}
+ \,
\frac{1}{5k^4} \hspace{5mm} \left( 24 (k \cdot p)^2 + 7 k^2 p^2 \right)
\hspace{3mm} \Bigg] \Bigg\}  {\rm ln} \frac{k^2}{p^2}  \\
\nonumber
\nonumber
\tau_3(k^2,p^2,q^2)&=& \hspace{3mm} \frac{\alpha}{12 \pi k^2} \Bigg\{ \hspace{5mm}
1 + \, \, 0 \, \, \frac{k
\cdot p}{k^2} - \, \frac{1}{5k^4} \hspace{5mm} \left( 4 (k \cdot p)^2 + 7 k^2 p^2
\right) \\
\nonumber
&& \hspace{43.5mm} - \frac{k \cdot p}{k^6} \hspace{5.3mm} \left( 2(k \cdot p)^2 + 3 k^2 p^2
\right) \\
\nonumber
&& \hspace{15mm} + \xi \Bigg[ 1 + \, \, \frac{3}{2} \, \, \frac{k \cdot p}{k^2}
+
\frac{1}{5k^4} \hspace{5mm} \left( 12 (k \cdot p)^2 +  k^2 p^2 \right) \\
\nonumber
&& \hspace{44mm} + 4 \frac{(k \cdot p)^3}{k^6} \hspace{37mm}
 \Bigg] \Bigg\}  {\rm ln} \frac{k^2}{p^2}  \\
\nonumber
\tau_6(k^2,p^2,q^2)&=&  \frac{\alpha (\xi-2)}{24 \pi k^2} \Bigg\{ \hspace{5mm}
1 + \, \hspace{2mm} \frac{k
\cdot p}{k^2} \hspace{2mm} + \, \frac{3}{5k^4} \hspace{5mm} \left( 2 (k \cdot p)^2 + k^2 p^2
\right) \\
\nonumber
&& \hspace{44mm} + \frac{4k \cdot p}{5k^6} \hspace{3mm} \left( 2(k \cdot p)^2 + k^2 p^2
\right) \hspace{8mm}  \Bigg\} {\rm ln} \frac{k^2}{p^2}  \\
\nonumber
\tau_8(k^2,p^2,q^2)&=& \hspace{0.5mm} - \frac{\alpha}{4 \pi k^2}
\hspace{1.5mm} \Bigg\{ \hspace{5mm}
1 + \, \hspace{0.7mm} \frac{2}{3}\frac{k
\cdot p}{k^2} \hspace{1mm} +  \frac{2}{3k^4} \, (k \cdot p)^2
\hspace{33mm} \Bigg\} {\rm ln} \frac{k^2}{p^2} \; .  \\ \label{eq:limit}
\end{eqnarray}
We learn the following points from the above calculation~:
\begin{itemize}

\item

To the lowest order in $1/k^2$, all the four $\tau_i$ are
independent of the angle between the momenta $k$ and $p$.

\item

Substituting these $\tau_i$ in the
expression for the full transverse vertex, we retrieve the
perturbative result for the transverse vertex derived
by Curtis and
Pennington~\cite{CP}, Eq.~(\ref{eq:largek}). This serves as one of the checks of the calculation.

\item

Comparing the equations for the real $\tau_i$,
Eqs.~(\ref{eq:Ayse2}-\ref{eq:Ayse8}), with their
large $k^2$ limit, Eq.~(\ref{eq:limit}), one can see
that all the $\Delta^2$ factors have
disappeared from the denominator. Hence, for large $k^2$, the $\tau_i$
are explicitly finite for all values of the angular variable.

\end{itemize}

 We can now use Eq. (\ref{eq:REAE}) to find out the large $k^2$
expansion of
the effective $\tau_{i}$. This yields~:
\vspace{2mm}
\begin{eqnarray}
\nonumber
\tau_{2}^{\rm eff}(k^2,p^2)&=&-\frac{\alpha}{12 \pi k^4} \; \,
\left\{\; 1 \;  - \; 2 \xi \; +
\frac{16}{5} \left( \; \; \frac{1}{3} \; \; - \; \; \, \, \xi \; \, \;
\right) \frac{p^2}{k^2} \right\}
\; {\rm ln} \frac{k^2}{p^2} \\ \nonumber
&& \; \\
\nonumber
\nonumber
\tau_{3}^{\rm eff}(k^2,p^2)&=& +\frac{\alpha}{12 \pi k^2} \; \, \left\{
\; 1 \;+
\; \frac{1}{4} \xi \,+ \; \frac{1}{5} \, \left( \; \; \frac{7}{3} \; \; -
\; \; \frac{3}{4} \xi \; \,
\right) \frac{p^2}{k^2} \right\} \; {\rm ln} \frac{k^2}{p^2} \\ \nonumber
&& \\
\nonumber
\nonumber
\tau_{6}^{\rm eff}(k^2,p^2)&=&\frac{\alpha (\xi-2)}{16 \pi k^2}
\; \,\left\{1 + \frac{5}{3} \frac{p^2}{k^2} \right\} \; {\rm ln} \frac{k^2}{p^2} \\  \nonumber
&&  \\
\nonumber
\nonumber
\tau_{8}^{\rm eff}(k^2,p^2)&=&- \frac{\alpha}{4 \pi k^2} \; \; \; \, \left\{ 1 +
\frac{1}{3} \frac{p^2}{k^2} \right\}  {\rm ln} \frac{k^2}{p^2}
 \\ \nonumber
&& \\
\nonumber
 \nonumber
\overline{\tau}_{\rm{eff}}(k^2,p^2)&=&-\frac{\alpha}{8 \pi k^2} \; \; \; \, \left\{
\; 1 +
\; \, \frac{1}{2} \xi \; - \, \frac{1}{3} \, \,\left(
\; \; 1 \; - \; \frac{11}{2} \xi \; \, \right)
\frac{p^2}{k^2} \right\} \; {\rm ln} \frac{k^2}{p^2} \quad . \\
\end{eqnarray}
Using the definition, Eq.~(\ref{eq:DW1}), for $U_1(x)$, we then deduce
its leading logarithmic form to be simply~:
\begin{eqnarray}
U_1(x)& \stackrel{x\to 0}{=} &{\alpha\over{2\pi}}\, {\rm ln}\ x \qquad .
\end{eqnarray}
The above equation is the scheme-independent perturbative expression for
$U_1(x)$ for $x \rightarrow 0$, to which every non-perturbative
construction must reduce in the weak coupling
regime. This is the main and remarkably simple result of this section.

\noindent Note firstly and importantly, all terms of the type ${\rm ln}x/x$ in the  equations for  the 
$\tau_i$
{\bf neatly} cancel out in the expression for $U_1(x)$. If this had not happened,
such terms would have led to non-integrable contributions.
 This cancellation is consistent with Eq.~(23),
which shows that there can be no  ${\rm ln}x/x$ term to
$0(\alpha)$. 
 Note secondly that the leading logarithmic perturbative expression for $U_1(x)$ turns out to be
independent of the gauge parameter! While we could imagine checking that 
$ \int_0^1 \, dx \,U_1(x)={4\pi} (\gamma - \nu)/{\alpha}$
numerically by constructing the integrand explicitly from Eqs.~(\ref{eq:DW1},
\ref{eq:REAE}-\ref{eq:Ayse8}) 
the lack of consistency
arising from the use of two different schemes would render
such an attempt meaningless (beyond leading logarithms).   

\noindent Importantly, our results are in agreement with the rules of the Landau-Khalatnikov
transformation~\cite{LK}.  These determine the gauge dependence of a Green's
function, once one knows its behaviour in some covariant gauge.
Thus, if in the Landau gauge
\begin{eqnarray*} 
F(p^2, \Lambda^2) \,=\,A_0 (p^2/\Lambda^2)^{\gamma_0}\, ,
\end{eqnarray*}
then these rules~\cite{LK,Zumino,Burden,Bashir} applied to quenched QED require that in a 
general covariant gauge 
\begin{eqnarray*}
F(p^2, \Lambda^2) \,=\,A (p^2/\Lambda^2)^{\gamma}\, ,
\end{eqnarray*}
where $\gamma\,=\,\gamma_0 +\,\alpha\xi/{4\pi}$ and $A, A_0$ are constants.
Thus, $\nu\,=\,\alpha\xi/(4\pi)$ provides the only gauge
dependence to the anomalous dimension~\cite{Zumino}-\cite{Bashir}.
Consequently, in Eqs.~(13-15,20-23,25,26), the factor $\gamma - \nu\,=\,\gamma_0$ is
gauge independent and in perturbation theory of ${\cal O}(\alpha^2)$.
Thus, $\int^1_0\,dx\ U_1(x)$ too must be of ${\cal O}(\alpha)$ and generally gauge
independent, like its $x\to 0$ limit, Eq.~(38).

\noindent Because the derivative condition, Eq.~(25), 
is merely a statement of the transverse vertex being
free of kinematic singularities, regardless of in what scheme
it has been calculated, it can be checked numerically. To ${\cal O}(\alpha)$, the derivative
condition reads~:
\vspace{2mm}
\begin{eqnarray}
  \omega \equiv  U_{1}(1)\; +  \; U_{1}^{\prime}(1)\,-
  \,\frac{16\pi}{\alpha}\ \gamma_0
= - \frac{3 \alpha\xi}{2 \pi}  \quad.\label{eq:WD}
\end{eqnarray}
Making use of the complete expressions in Eqs.~(\ref{eq:DW1},
\ref{eq:REAE}-\ref{eq:Ayse8}),
we plot  $\omega/ \alpha$ versus the
gauge parameter $\xi$ in Fig 2. The
numerical and analytical results
are in excellent agreement with each other.

\newpage

\section{Conclusions}

The non-perturbative study of the fermion propagator through
its Schwinger-Dyson equation requires an {\em ansatz} for
the fermion-gauge boson vertex.
 Here, we have shown that this
vertex (in the case of massless fermions) can be expressed in
terms of a single unknown function $U_1(x)$ constrained to ensure the 
multiplicative
renormalizability of the fermion propagator. 
We have devised a general non-perturbative
form for this function and so developed a simple construction for the full
fermion-boson vertex. We have then calculated its perturbative expansion and
found the remarkably simple result that to $O(\alpha)$~:
\begin{eqnarray*}
U_1(x)& \stackrel{x\to 0}{=} &{\alpha\over{2\pi}}\, {\rm ln}\ x \qquad .
\end{eqnarray*} 
Any non-perturbative {\it ansatz} for $U_1(x)$ should agree with this in the weak coupling limit.
This should help in pinning down the only unknown part of the full interaction
vertex, Eqs.~(5,6,8,13-15), and so finally encapsulate the physics encoded in the Schwinger-Dyson
equation for the fermion propagator. \\
\vskip 2cm

\noindent
{\bf Acknowledgements} \\
AB wishes to thank the International Centre
for Theoretical Physics (ICTP), Trieste, for providing the funding
for his post doctoral research. AB and AK are grateful for the hospitality
offered to them by the ICTP for their stay there in the summer of 1996.
We are grateful to John Gracey for  helpful comments on the fermion anomalous dimension.

\vfil\eject
\hsize=16.5cm

\vfil\eject
\vskip 1cm
\noindent{\bf FIGURES}
\begin{figure}[h]
\begin{center}
\mbox{~\epsfig{file=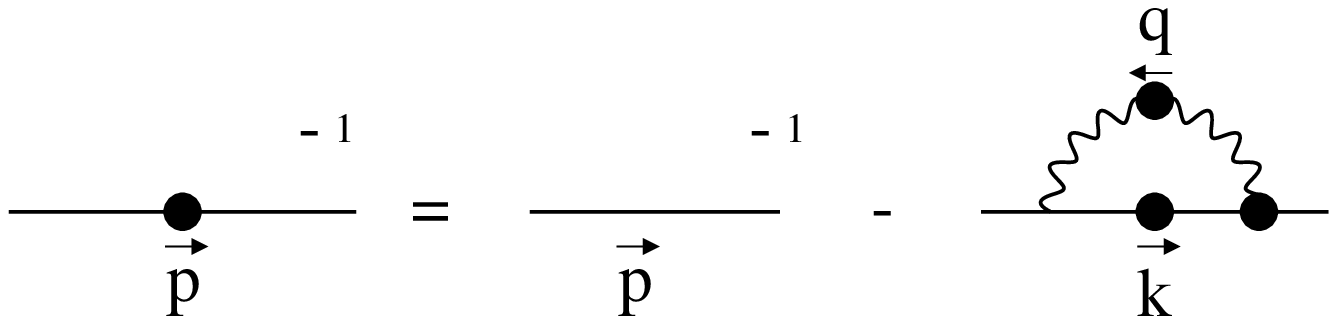,width=400pt}}
\caption{Schwinger-Dyson equation  for the fermion propagator.
The straight lines represent fermions and the wavy line
the photon.  The solid dots
indicate full as opposed to
bare, quantities.}
\end{center}
\end{figure}
\vspace{-1.5cm}
\begin{figure}[h]
\begin{center}
\mbox{~\epsfig{file=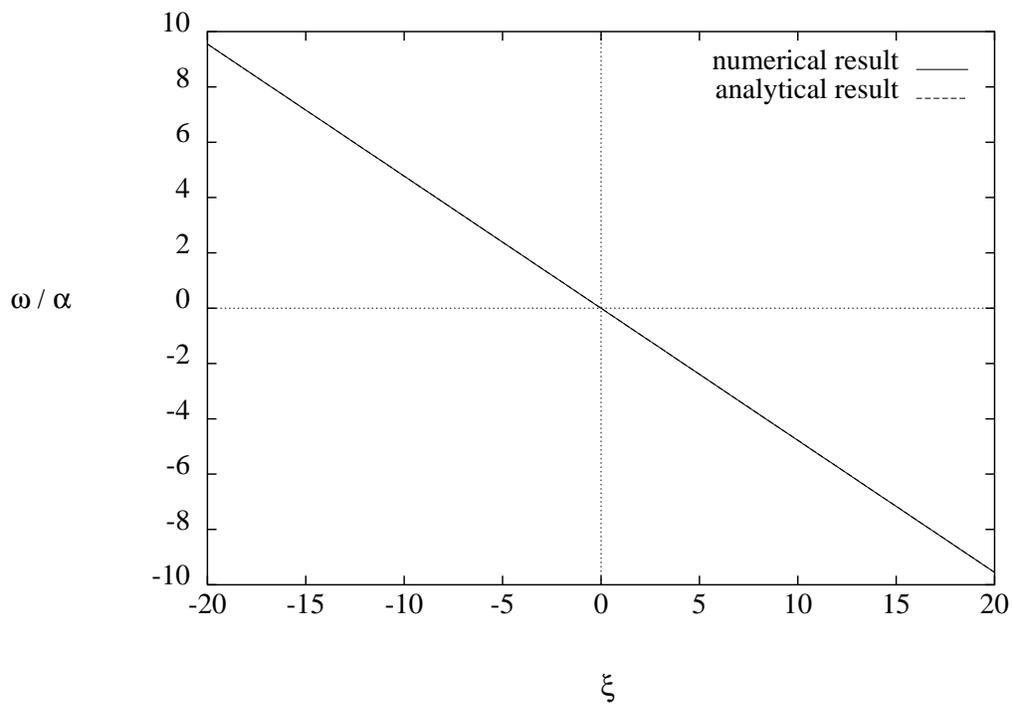,angle=0,width=250pt}}
\vspace{-1.2cm}
\caption{ $\omega/\alpha$ of  Eq.~(\protect\ref{eq:WD}), 
 is plotted as 
a function of the gauge parameter $\;\xi\;\,$.  The solid line which
represents the numerical result lies
completely on top of the
dashed analytical result, $\, -3\xi/{4\pi}\, $
 of  Eq.~(\protect\ref{eq:WD}), in perfect agreement.}
\end{center}
\end{figure}

\end{document}